\documentclass{aa}
\usepackage{graphicx}
\usepackage{natbib}
\usepackage{txfonts}
\usepackage{float}
\bibpunct{(}{)}{;}{a}{}{,}    % to follow the A&A style
\usepackage{cancel}
\usepackage{xcolor}
\usepackage[colorlinks=true,citecolor=blue,linkcolor=blue]{hyperref}
\usepackage{soul} % package for strikethrough \st{}

\def\apjs{ApJS}

\def\aap{Astron. Astrophys.}

\begin{document}

  % \title{Solar rotational variability in different stellar filter systems}
  \title{Connecting measurements of solar and stellar brightness variations}

   \author{N.-E. Nèmec,
          \inst{1}
          E. I\c{s}{\i}k\inst{2,1},
          A. I. Shapiro\inst{1},
          S. K. Solanki\inst{1,3},
          N. A. Krivova\inst{1}
          and Y. Unruh\inst{4}
              }

   \institute{Max-Planck-Institut für Sonnensystemforschung, Justus-von-Liebig-Weg 3, D-37077 Göttingen, Germany\\
              \email{nemec@mps.mpg.de} 
             \and 
             Dept. of Computer Science, Turkish-German University, \c{S}ahinkaya Cd. 108, 34820 Beykoz, Istanbul, Turkey
            \and 
            School of Space Research, Kyung Hee University, Yongin, Gyeonggi, 446-701, Korea
            \and Imperial College, Astrophysics Group, Blackett Laboratory, London SW7 2BZ, United Kingdom
             }

   \date{}
                                            
% \abstract{}{}{}{}{} 
% 5 {} token are mandatory
 
  \abstract
 {Comparing solar and stellar brightness variations is hampered by the difference in spectral passbands used in observations as well as by the possible difference in the inclination of their rotation axes from the line of sight.}
   {We calculate the rotational variability of the Sun as it would be measured in passbands used for stellar observations.
   In particular, we consider the filter systems used by the CoRoT, \textit{Kepler}, TESS, and \textit{Gaia} space missions. We also quantify the effect of the inclination of the rotation axis on the solar rotational variability.}
   {We employ the Spectral And Total Irradiance REconstructions (SATIRE) model to calculate solar brightness variations in different filter systems as observed from the ecliptic plane. 
  We then combine the simulations of the surface distribution of the magnetic features at different inclinations using a surface flux transport model (SFTM) with the SATIRE calculations to
compute the dependence of the variability on the inclination. }
  {For an ecliptic-bound observer, the amplitude of the solar rotational variability, as observed in the total solar irradiance (TSI) is 0.68 mmag (averaged over solar cycles 21-24). We obtained corresponding amplitudes in the
 \textit{Kepler} (0.74 mmag), CoRoT (0.73 mmag), TESS (0.62 mmag), \textit{Gaia G}  (0.74 mmag),  \textit{Gaia G$_{RP}$}  (0.62 mmag), and \textit{Gaia G$_{BP}$} (0.86 mmag) passbands. Decreasing the inclination of the rotation axis decreases the rotational variability. For a sample of randomly inclined stars, the variability is on average 15\% lower in all filter systems considered in this work. This almost compensates for the difference in the amplitudes of the variability in TSI and \textit{Kepler} passbands, making the amplitudes derived from the TSI records an  ideal representation of the solar rotational variability for comparison to \textit{Kepler} stars with unknown inclinations. }
  {The TSI appears to be a relatively good measure of solar variability for comparing with stellar measurements in the CoRoT, \textit{Kepler},  TESS \textit{Gaia G}, and \textit{Gaia G$_{RP}$} filters.  Whereas the correction factors can be used to convert the amplitude of variability from solar measurements into the values expected for stellar missions, the inclination affects the shapes  of the light curves so that a much more sophisticated correction than simple scaling is needed to obtain light curves out of ecliptic for the Sun.}


%\titlerunning{Solar rotational variability in different filter systems}
\titlerunning{Connecting solar and stellar variabilities}
\authorrunning{Nèmec et al.}
\maketitle
%
%-------------------------------------------------------------------

\section{Introduction}
The dedicated planet-hunting photometric missions such as  CoRoT \citep[Convection, Rotation and planetary Transit, see][]{COROT,COROT2}, \textit{Kepler} \citep{KEPLER}, and TESS \citep[Transiting Exoplanet Survey Satellite, see][]{TESS}, as well as the \textit{Gaia} space observatory \citep{Gaia2016} have made it possible to measure stellar brightness variability with unprecedented precision. In particular, they allow studying stellar brightness variations caused by transits (as the star rotates) and the evolution of magnetic features, i.e. bright faculae and dark spots. Such variations are often referred to as rotational stellar variability. 
The plethora of stellar observational data rekindled an interest in the questions of how typical is our Sun as an active star and, more specifically, how does the solar rotational variability compare to that of solar-like stars? Furthermore, these data allow
probing if the solar activity paradigm is valid for other stars as well. This requires comparing the stellar properties and behaviour with those of the Sun.  
While the solar variability has been measured for over four decades now by various dedicated space missions  \citep[see, e.g.,][for reviews]{Froehlich2012,Ermolli2013,Solanki2013, Greg2016},
a comparison between solar and stellar brightness measurements is far from straightforward \citep[see, e.g.,][]{Basri2010,Timo2020,Witzke2020}. Firstly, solar and stellar brightness variations  have been measured in different spectral passbands. Since the amplitude of the solar rotational variability strongly depends on the wavelength \citep{Solanki2013,Ermolli2013}, the solar and stellar brightness records can be reliably compared only after conversion from one passband to another. Secondly, the solar brightness variations have (so far) only been measured from the ecliptic plane  which is very close to the solar equatorial plane (the angle between the solar equator and ecliptic plane is about 7.25$^\circ$). The values of the angle between the line of sight of the observer and the rotation axes of the observed stars (hereinafter referred to as the inclination) are mostly unknown. 

Studies comparing solar and stellar rotational brightness variations have used different types of solar brightness measurements. \cite{Timo2020}, for instance, used the total solar irradiance (TSI), i.e. the solar radiative flux at 1~AU integrated over all wavelengths. More commonly, however, the solar variability was characterised \citep[see, e.g.,][]{Basri2010,Gilliland2011,Harrison2012} using measurements  by the Variability of solar IRradiance and Gravity Oscillations / Sun PhotoMeters (VIRGO/SPM) \citep[][]{Froehlich1995,Froehlich1997} instrument on-board the Solar and Heliospheric Observatory (SoHO). VIRGO/SPM measures solar brightness in three filters with a bandwidth of 5 nm each. Neither VIRGO/SPM nor TSI measurements can be directly compared to records of stellar brightness variability, which typically cover wavelength ranges broader than the VIRGO/SPM filters, but much narrower than the TSI. Therefore accurate estimations of solar variability in passbands used for stellar measurements has so far been missing.
Some effort has been already put in the modelling of the solar rotational variability as it would be observed out-of-ecliptic \citep[e.g.,][]{Vieira2012,Shapiro2016,Nina2020}. In particular, 
\cite{Shapiro2016} and \cite{Nina2020} (hereinafter N20) have shown that the amplitude of the solar brightness variations on the rotational timescale decreases with decreasing inclination. Consequently, due to its almost equator-on view, the Sun would appear on average more variable than stars with the same activity level but observed at random inclinations. At the same time, an easy to use receipt for correcting variability for the inclination effect  has been missing until now and, consequently,  inclination has not yet been quantitatively accounted for in solar-stellar comparison studies.

In this paper we seek to overcome both hurdles and quantify solar variability in passbands used by different stellar space missions and at different inclinations. In Sect.~\ref{Equator} we employ the Spectral And Total Irradiance REconstruction \citep[SATIRE;][]{Fligge2000,Krivova2003} model of solar brightness variations to show how the actual solar brightness variations relate to solar brightness variations as they would be observed in spectral passbands used by stellar missions.  We also establish the connection between the TSI and VIRGO/SPM measurements. In Sect. \ref{Inclination} we follow the approach developed by N20 to quantify the effect of the inclination on the brightness variations. We discuss, how the Sun as observed by \textit{Kepler} can be modelled using light curves obtained by VIRGO/SPM in Sect. \ref{Kepler-VIRGO} 
before we summarise our results and draw conclusions in Sect.~\ref{Conclusion}.

\section{Conversion from solar to stellar passbands}\label{Equator}

\subsection{SATIRE-S}
The Spectral And Total Irradiance REconstruction \cite[SATIRE,][]{Fligge2000,Krivova2003} model attributes the brightness variations of the Sun on timescales longer than a day to the presence of magnetic features, such as bright faculae and dark spots, on its surface. The two main building blocks of SATIRE are the areas and the positions of the magnetic features on the solar disc as well as contrasts of these features relative to the quiet Sun (i.e. regions on the solar surface free from any apparent manifestations of magnetic activity).
The contrasts of the magnetic features as a function of disc position and wavelength  were computed
by \cite{Unruh1999} with the spectral synthesis block of the  ATLAS9 code \citep{Kurucz1992, Castelli1994}. The 1D atmospheric structure of the two spot components (umbra and penumbra) and of the quiet Sun were calculated using radiative equilibrium models produced with the ATLAS9 code, while the facular model is a modified version of FAL-P by \cite{Fontenla1993}.

Various versions of the SATIRE model exist. In this section we employ the most precise version, which is SATIRE-S, where the suffix ``S'' stands for the satellite era \cite[][]{Ball2014,Yeo2014}.  SATIRE-S uses the distribution of magnetic features on the solar disc obtained from observed magnetograms and continuum disc images and spans from 1974 to today, covering four solar cycles. As especially the early ground-based  observations contain gaps in the data we use the SATIRE-S model as presented by \cite{Yeo2014} (version 20190621), where the gaps in Spectral Solar Irradiance (SSI) and TSI have been filled using the information provided by solar activity indices. SATIRE-S was shown to reproduce the apparent variability of the Sun as observed, in both the SSI and in the TSI \cite[see][and references therein]{Ball2012,Ball2014,Yeo2014,Danilovic2016}.
The spectral resolution of the SATIRE output is 1 nm below 290 nm, 2 nm between 290 nm and 999 nm and 5 nm above 1000 nm. This is fully sufficient for the calculations presented in this study.

\subsection{Filter systems}\label{Filters}
In this section we multiply the SATIRE-S SSI output with  the response function of a given filter and integrate it over the entire filter passband to obtain the solar light curve in the corresponding filter. It is important to take the nature of the detectors used in different instruments into account \citep[see e.g.][]{Maxted2018}.
In particular, while solar instruments (e.g. VIRGO/SPM and all TSI instruments) measure the energy of the incoming radiation, charge-coupled devices (CCDs) utilised in \textit{Kepler},  \textit{Gaia}, and TESS count the number of photons  and not their energy. In order to obtain the solar light curve, $LC$, as it would be measured  by the instrument counting photons we therefore follow
\begin{equation}
LC =  \int\limits_{\lambda_1}^{\lambda_2} R(\lambda) \cdot I(\lambda)  \frac{\lambda}{h\cdot c} \, d\lambda,
\label{eq:filter}
\end{equation}
\noindent where $\lambda_1$ and $\lambda_2$ are the blue and red threshold wavelengths of the filter passband, $R(\lambda)$ is the response function of the filter and $I(\lambda)$ is the spectral irradiance at a given wavelength, $h$ is the Planck constant, and $c$ the speed of light.

\begin{figure}
\centering
\includegraphics[width=\columnwidth]{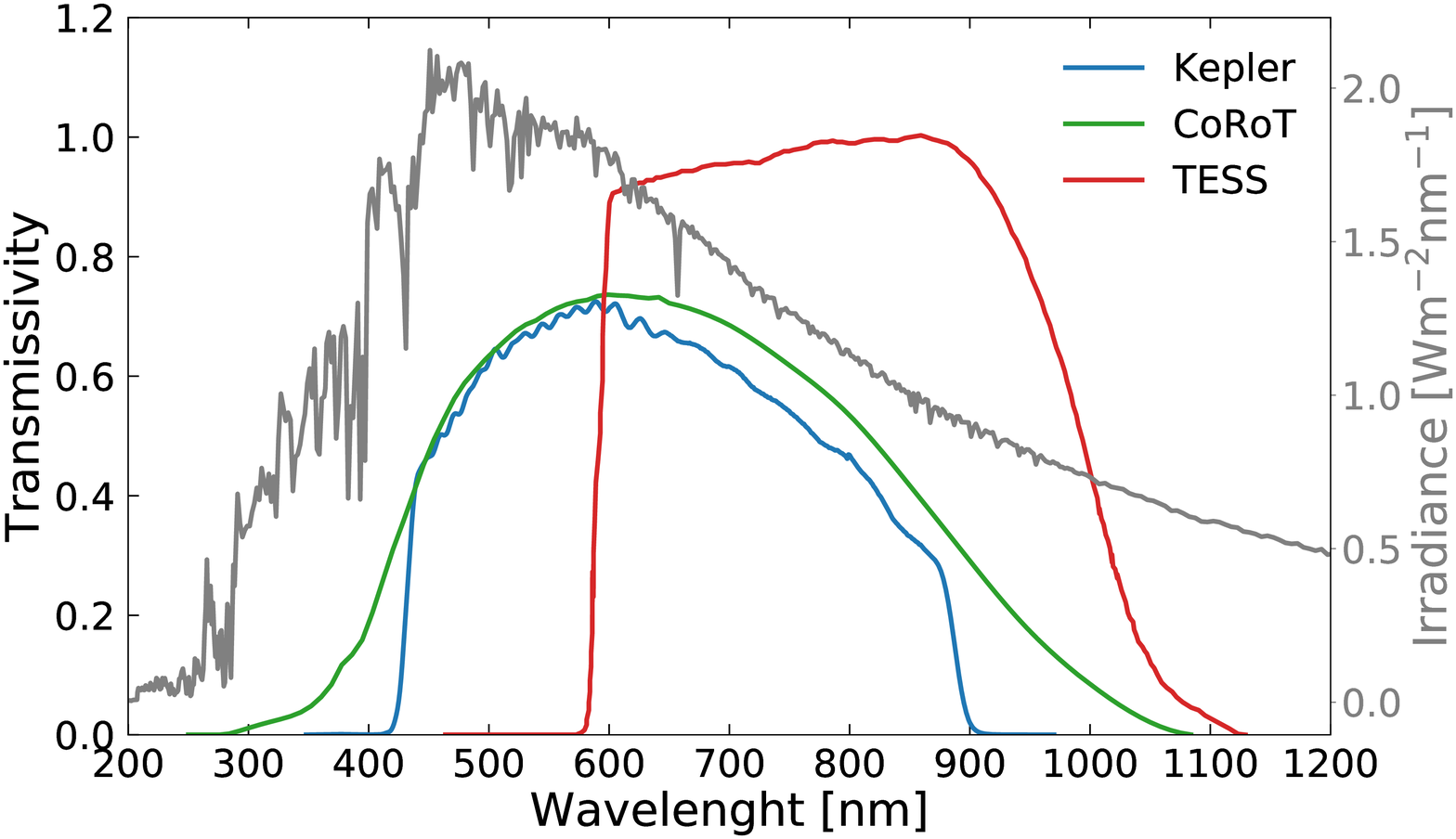}
\includegraphics[width=\columnwidth]{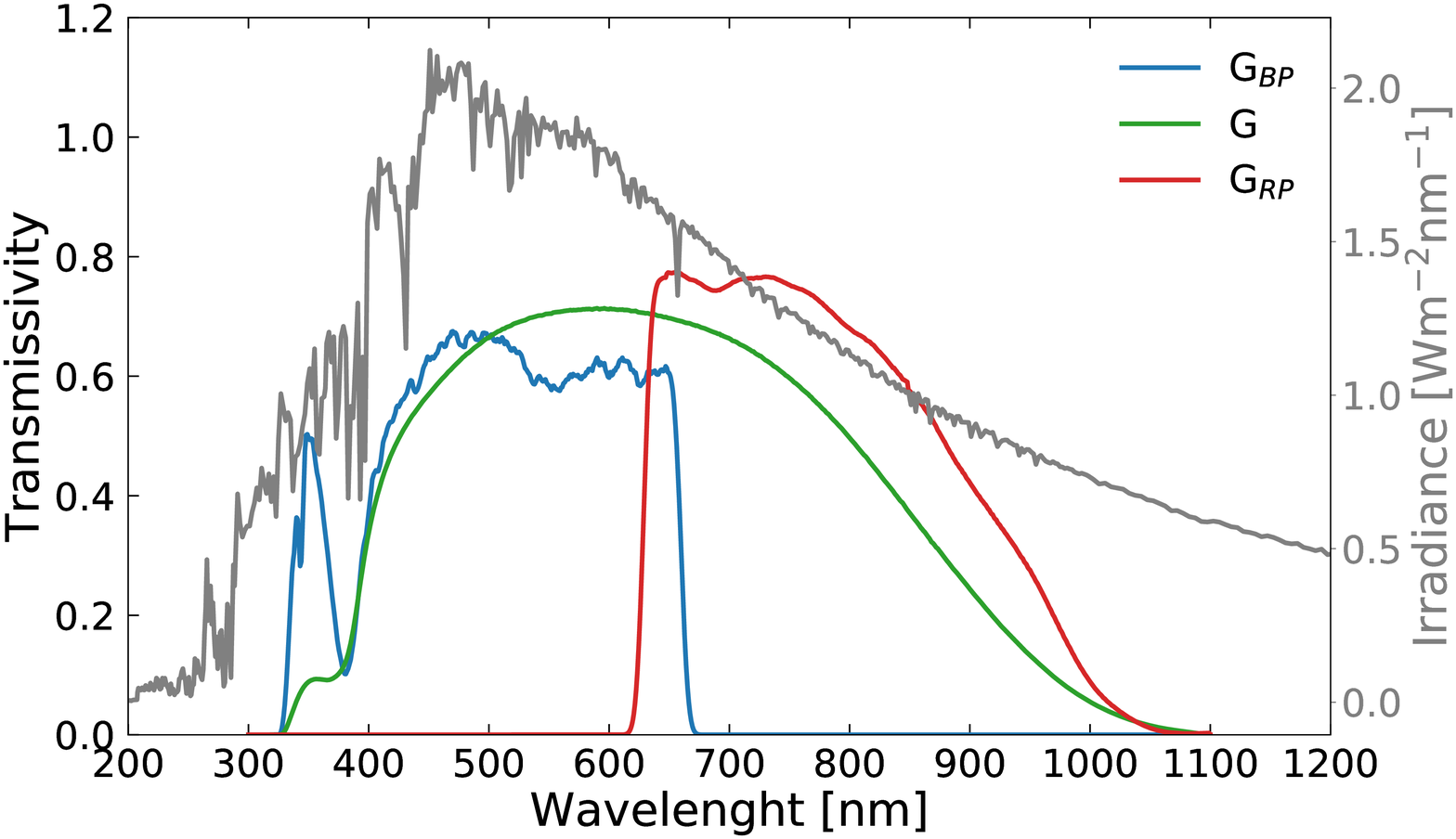}
\includegraphics[width=\columnwidth]{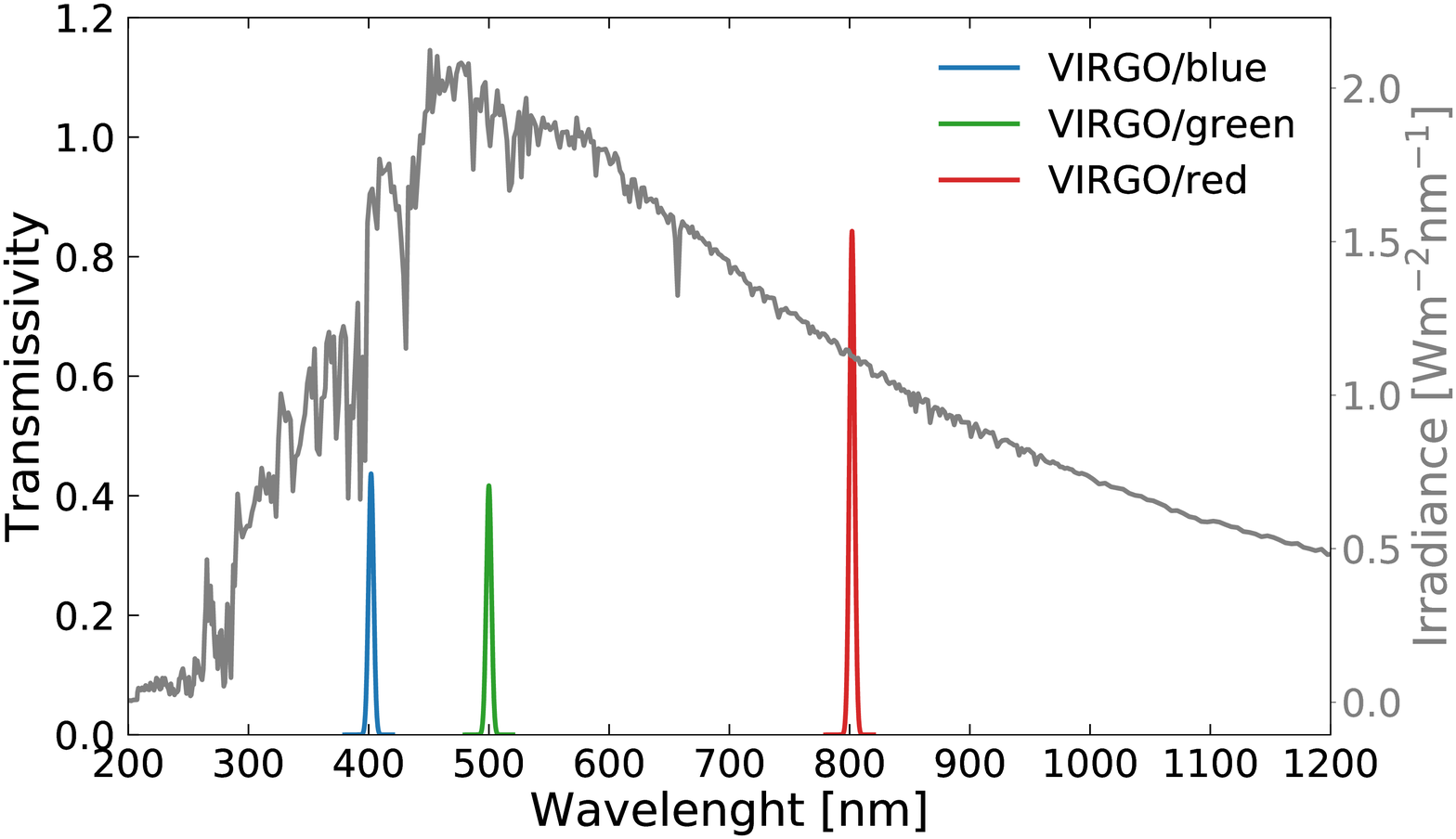}
\caption{Response functions for the various filter systems used in this study. For comparison, the quiet-Sun spectrum used by SATIRE-S is plotted in grey in each panel. Top panel: \textit{Kepler}, TESS and CoRoT; middle panel: the
three Gaia passbands; bottom panel: the three VIRGO/SPM channels.}
\label{fig:filters}
\end{figure}

First we consider several broad-band filters used by the planetary-hunting missions:  CoRoT, \textit{Kepler}, and TESS. The spectral passbands employed in these missions are shown in the top panel of Fig.~\ref{fig:filters}, along with the quiet-Sun spectrum calculated by \cite{Unruh1999} and used in SATIRE-S. Clearly, the  CoRoT and \textit{Kepler} response functions are very similar to each other, as both missions focused on G-stars. TESS is designed to observe cooler stars compared to \textit{Kepler}, hence the response function is shifted towards the red part of the spectrum.

\textit{Gaia} measures stellar brightness in three different channels \citep{Gaia2016}. \textit{Gaia G}  is sensitive to photons between 350 and 1000 nm. Additionally, two prisms disperse 
 the incoming light between 330 and 680 nm for the `Blue Photometer' (hereafter, referred to as \textit{Gaia G$_{BP}$}) and between 640--1050 nm for the `Red Photometer' (hereafter, referred to as \textit{Gaia G$_{RP}$}). The response functions are shown in the middle panel in Fig.~\ref{fig:filters}. We employ the revised passbands used for the second data release of Gaia \citep[Gaia DR2,][]{Evans2018} for the calculations.

Solar-stellar comparison studies have often used the solar variability as it is measured by the VIRGO/SPM instrument. SPM comprises three photometers, with a bandwidth of 5 nm operating at 402 nm (blue), 500 nm (green), and 862 nm (red). The response functions are shown in Fig.~\ref{fig:filters} in the bottom panel. We refer to these filters from now on simply as VIRGO/blue/green/red.

\subsection{Results} \label{results_eq}

\begin{figure*}
\centering
\includegraphics[width=\textwidth]{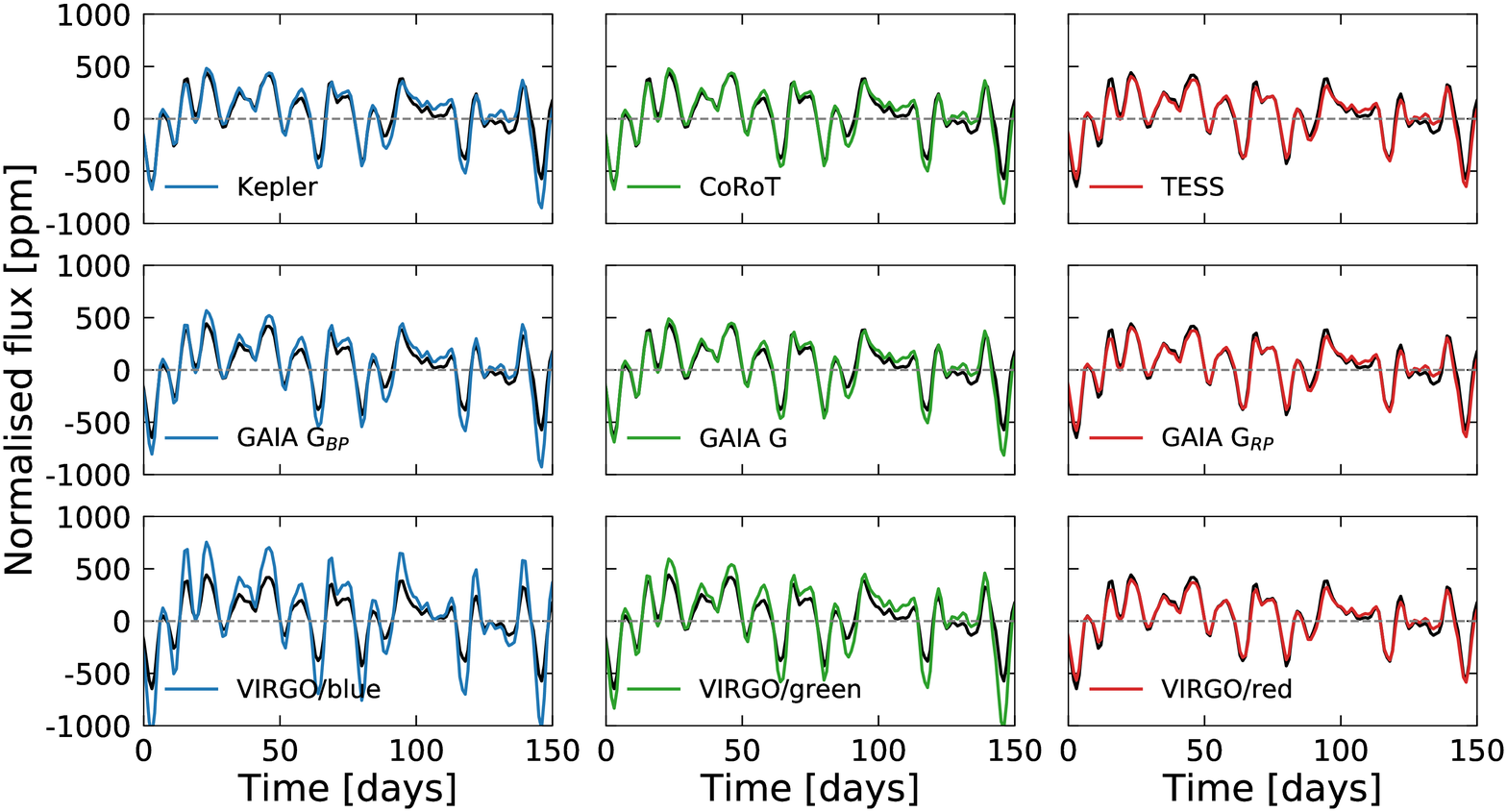}
\caption{Normalised fluxes for the different filter systems compared to the TSI (black solid line). Top panels: \textit{Kepler}, CoRoT and TESS, middle panels: the three \textit{Gaia} passbands, bottom panels : the three VIRGO/SPM channels.}
\label{fig:filters_LC}
\end{figure*}

\begin{figure*}
\centering
\includegraphics[width=\textwidth]{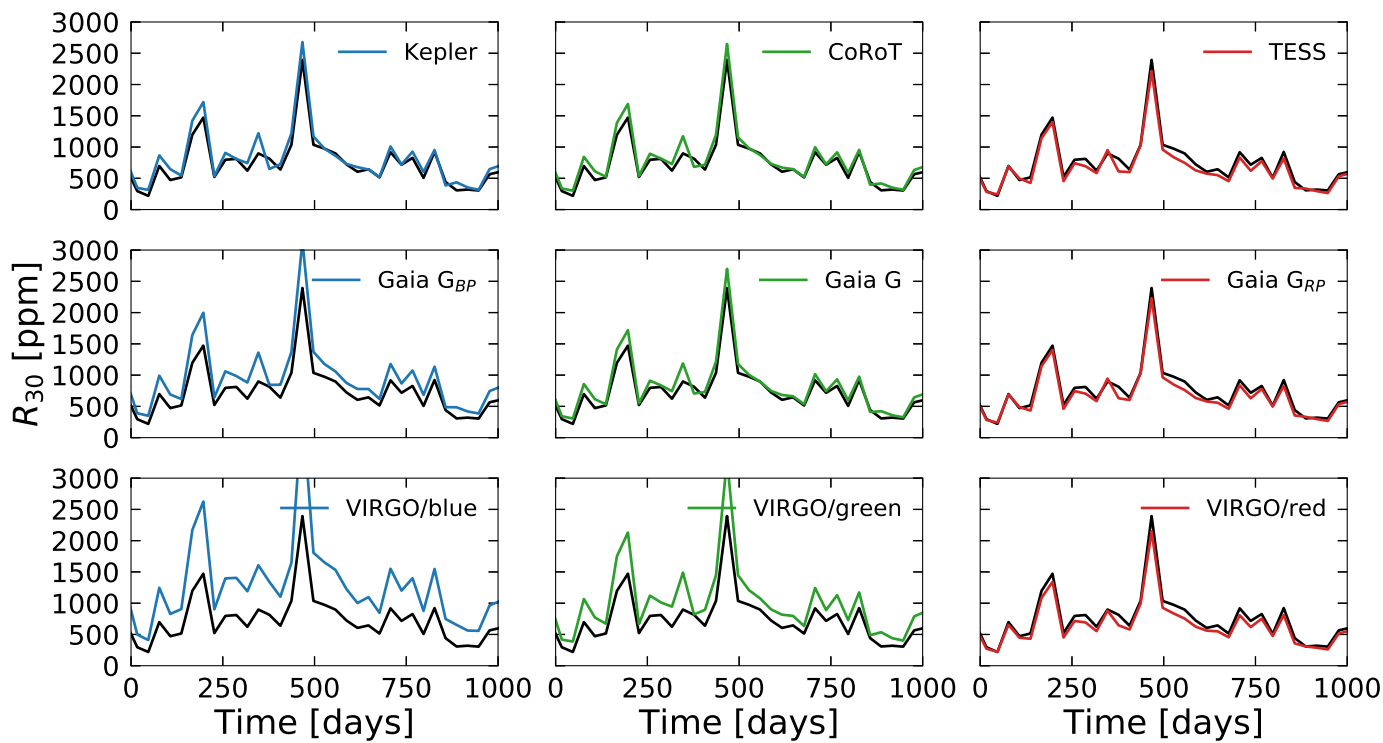}
\caption{$R_{30}$ in different filter systems compared to the TSI (black solid line) for a timespan of 1000 days over solar cycle 22. Top panels: \textit{Kepler}, CoRoT and TESS, middle panels: the three \textit{Gaia} passbands, bottom panels: the three VIRGO/SPM channels. See the main text for the definition of $R_{30}$.} 
\label{fig:filters_R_30}
\end{figure*}

\begin{figure*}
\centering
\includegraphics[width=\textwidth]{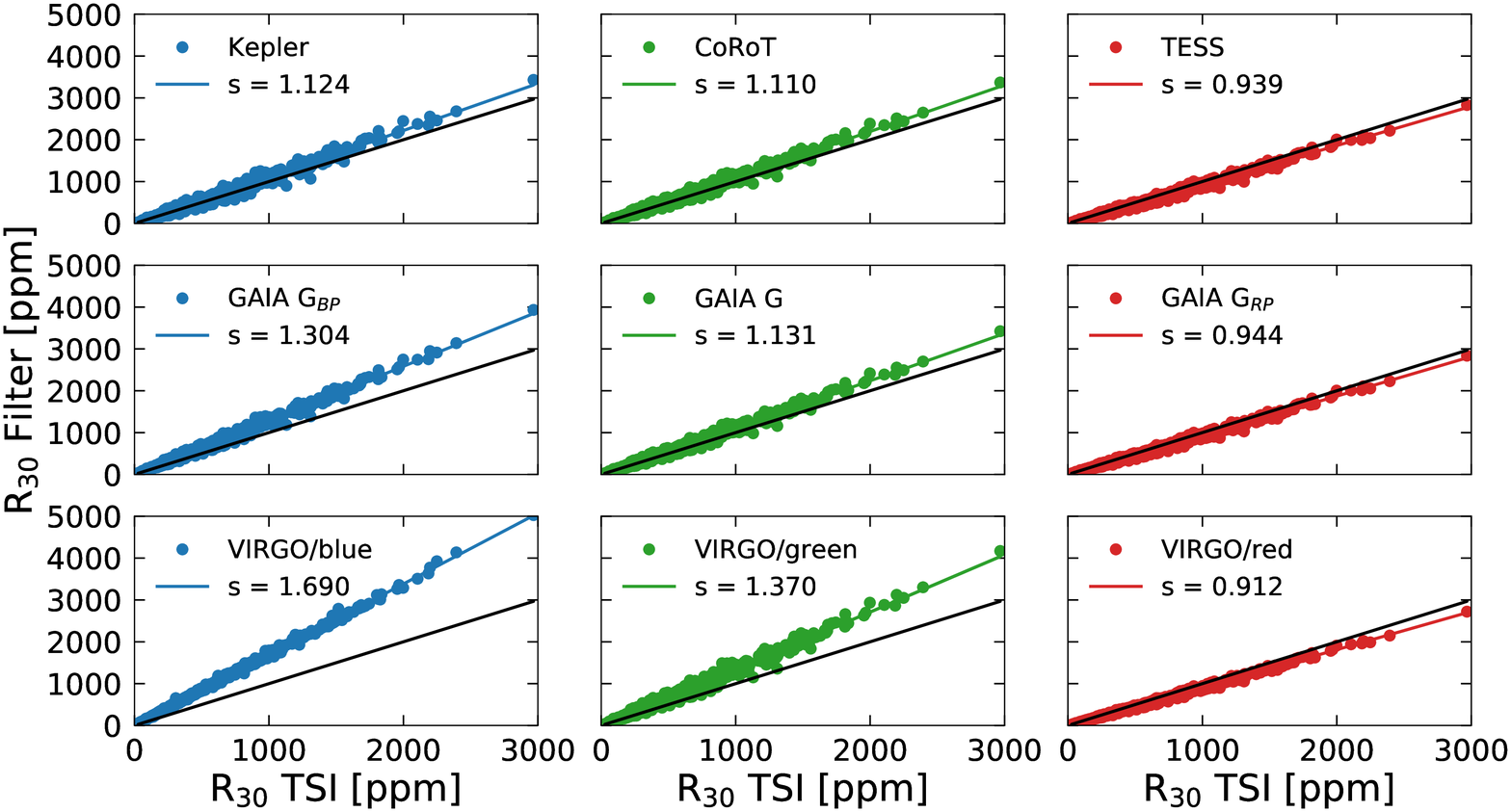}
\caption{Linear regression between the $R_{30}$-values calculated with the TSI and with the solar light curves as they would be recorded in different filter systems. The values of the slope, $s$, are given in the legend. The black lines have a slope equal to 1.}
\label{fig:R_30_regressions}
\end{figure*}

Figure~\ref{fig:filters_LC} shows the solar light curve for the period of 2456700 -- 2456850 JD (24 February 2014 -- 11 July 2014), as it would be observed in different passbands. This corresponds to a 150 --day interval during the maximum of cycle 24. This interval was chosen arbitrarily to display the effect of the filter systems on the solar variability. For this, we first divided each light curve in 90--day segments. This time-span corresponds to  \textit{Kepler}-quarters. This is motivated by the way \textit{Kepler} observations are gathered and reduced. We note, that the detrending by Kepler operational mode is applied here for purely illustrative purposes and it is not used for calculations presented below. Within each segment, we subtracted the mean value from the fluxes, before dividing the corresponding values by the mean flux in each segment. In all stellar broad-band filters, the light curve turns to be remarkably similar in shape to the TSI (solid black curve), although the amplitude can differ. As one can expect, the difference in the amplitude of the variability is somewhat more conspicuous in the blue filters. The $Gaia~G_{RP}$ light curve is basically identical to the TSI light curve, whereas the variability in $Gaia~G_{BP}$ and in the narrow VIRGO-blue filter show far stronger variability than the TSI.

To quantify the rotational variability, we compute the $R_{30}$ values \citep[see e.g.][]{Basri2013}. For this, the obtained light curves are split into 30-day segments and within each segment, we calculate the difference between the extrema, and divide this value  by the mean flux in the segment to get the relative variability. For the SATIRE-S time series we directly consider the difference between the extrema instead of the differences between the 95th and 5th percentiles of sorted flux values, as is usually done in the literature with the more noisy \textit{Kepler} measurements. We calculate $R_{30}$ values for the period  1974--2019 (i.e. cycles 22--24). This allows us to quantify the mean level of solar variability in $R_{30}$ that represents the full four decades of TSI measurements.

 \begin{table}
\begin{center}
\begin{tabular}{ll}
\hline \hline
		&	slope 	\\
\hline
\textit{Kepler}	       	    &	1.123 ($\pm$0.007) \\
CoRoT               		&	1.110 ($\pm$0.006) \\
TESS	                	&	0.939 ($\pm$0.004) \\
\textit{Gaia G$_{BP}$}		&	1.304 ($\pm$0.005) \\
\textit{Gaia G}	        	&	1.131 ($\pm$0.005) \\
\textit{Gaia G$_{RP}$}		&	0.944 ($\pm$0.004) \\
VIRGO/blue              	&	1.689 ($\pm$0.003) \\
VIRGO/green             	&	1.370 ($\pm$0.007) \\
VIRGO/red	           	    &	0.912 ($\pm$0.003) \\
\hline \hline
\end{tabular}
\end{center}
\caption{Slopes of the linear regressions in Fig. \ref{fig:R_30_regressions}.}
\label{tab:regressions}
\end{table}

 \begin{table}
\begin{center}
\begin{tabular}{llllll}
\hline \hline
	&	21	&	22	&	23	&	24	&	mean	\\
\hline
TSI	&	0.743	&	0.806	&	0.682	&	0.492	&	0.681	\\
\textit{Kepler}	&	0.808	&	0.872	&	0.731	&	0.530	&	0.735	\\
CoRoT	&	0.801	&	0.866	&	0.726	&	0.526	&	0.730	\\
TESS	&	0.680	&	0.739	&	0.615	&	0.445	&	0.620	\\
\textit{Gaia G$_{BP}$}	&	0.944	&	1.020	&	0.861	&	0.625	&	0.862	\\
\textit{Gaia G}	&	0.817	&	0.883	&	0.741	&	0.537	&	0.744	\\
\textit{Gaia G$_{RP}$}	&	0.684	&	0.742	&	0.617	&	0.447	&	0.623	\\
VIRGO/blue	&	1.252	&	1.352	&	1.167	&	0.846	&	1.154	\\
VIRGO/green	&	0.983	&	1.056	&	0.894	&	0.653	&	0.897	\\
VIRGO/red	&	0.665	&	0.722	&	0.600	&	0.435	&	0.606	\\
\hline \hline	
%\hline \hline
\end{tabular}
\end{center}
\caption{Cycle-averaged $R_{30}$ in mmag.}
\label{tab:mean_R_30}
\end{table}

In Fig.~\ref{fig:filters_R_30} we show the $R_{30}$ values for all the filter systems introduced in Sect.~\ref{Filters} in comparison to $R_{30}$ of the TSI for a 1000-day interval starting July 26 2013. This interval therefore includes the maximum of solar 24 as well.
To better quantify the dependence of $R_{30}$  on the passband, we show linear regressions between the variability $R_{30}$ in each filter system and the TSI in Fig.~\ref{fig:R_30_regressions}.
The slope of the linear regressions are listed in Table \ref{tab:regressions}. The Pearson-correlation coefficient is above 0.98 for all of the filter systems.
 Naturally, the slope of the linear regressions depends on the considered filter system. For example, TESS and \textit{Gaia~G} regressions have a slope close to 1, but $G_{BP}$ displays a slope \textgreater 1, whereas VIRGO-red exhibits  a slope \textless 1. Unsurprisingly, the slope is highest for the blue VIRGO filter, where the amplitude of the variability is highest.   We note that the good agreement of the TSI with the red filters is expected to be valid only for the rotational variability, which is dominated by spots. In contrast, the solar irradiance variability on the activity cycle timescale is given by the delicate balance between facular and spot components and, consequently, has a very sophisticated spectral profile \citep{Shapiro2016,Witzke2018}. Thus, values of slopes from Table~\ref{tab:regressions} can not be extrapolated from the rotational to the activity cycle timescales \citep[see][for the detailed discussion]{Shapiro2016}.

Table \ref{tab:mean_R_30} lists the cycle-averaged values of $R_{30}$ for all passbands in mmag. All in all, Fig.~\ref{fig:R_30_regressions}, and Tables \ref{tab:regressions} and \ref{tab:mean_R_30} show that the TSI is a passable representative for the variability on the solar rotation timescale as it would be observed in the TESS, \textit{Kepler}, CoRoT, \textit{Gaia G}, \textit{Gaia G$_{RP}$} and VIRGO-red filters, but it noticeably underestimates the variability in $G_{BP}$, VIRGO-green VIRGO-blue.

\begin{figure}
\centering
\includegraphics[width=0.50\textwidth]{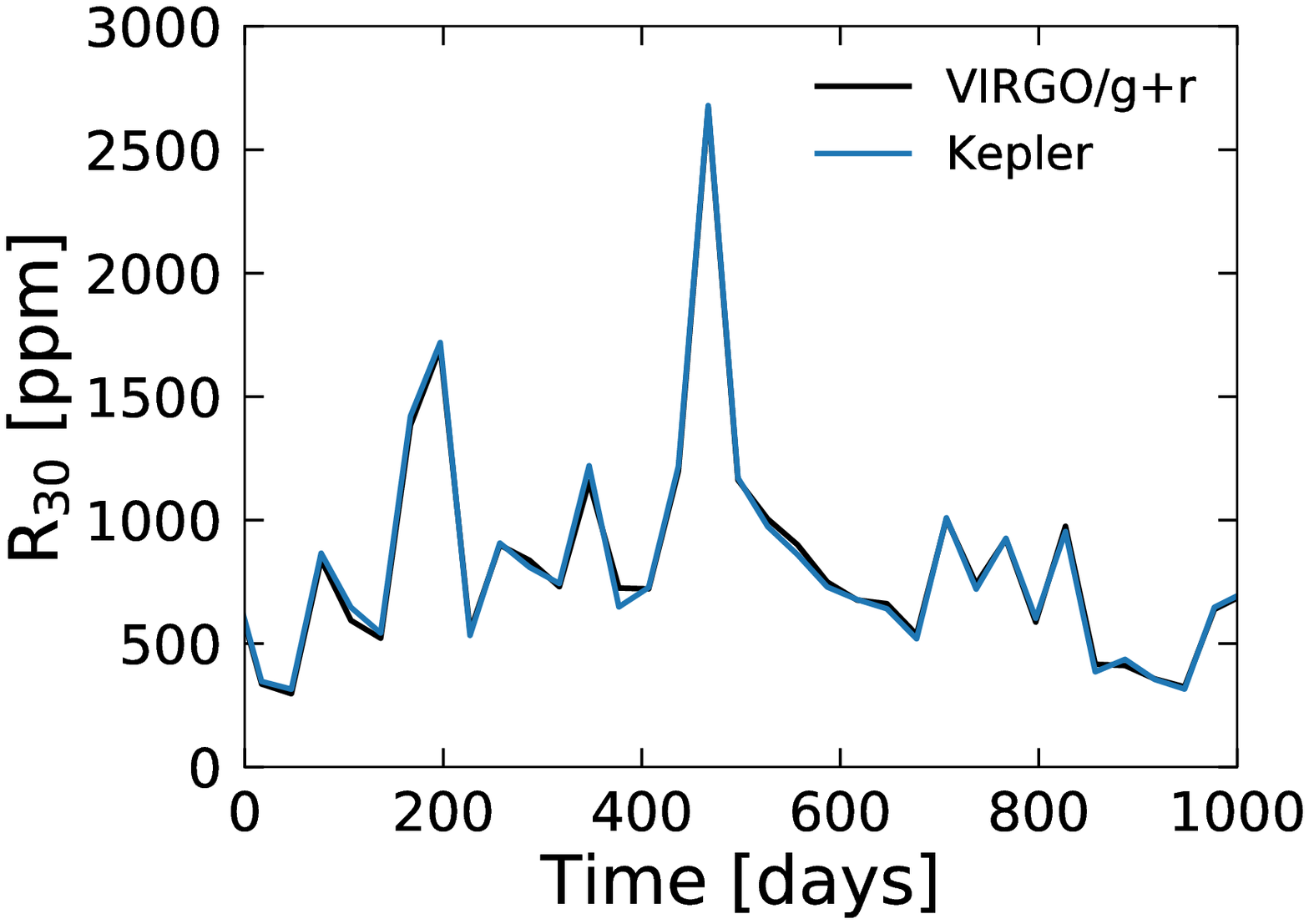}
\caption{$R_{30}$ over cycle 22 for VIRGO/g+r and \textit{Kepler}.} 
\label{fig:K_V_LC}
\end{figure}

\begin{figure}
\centering
\includegraphics[width=0.50\textwidth]{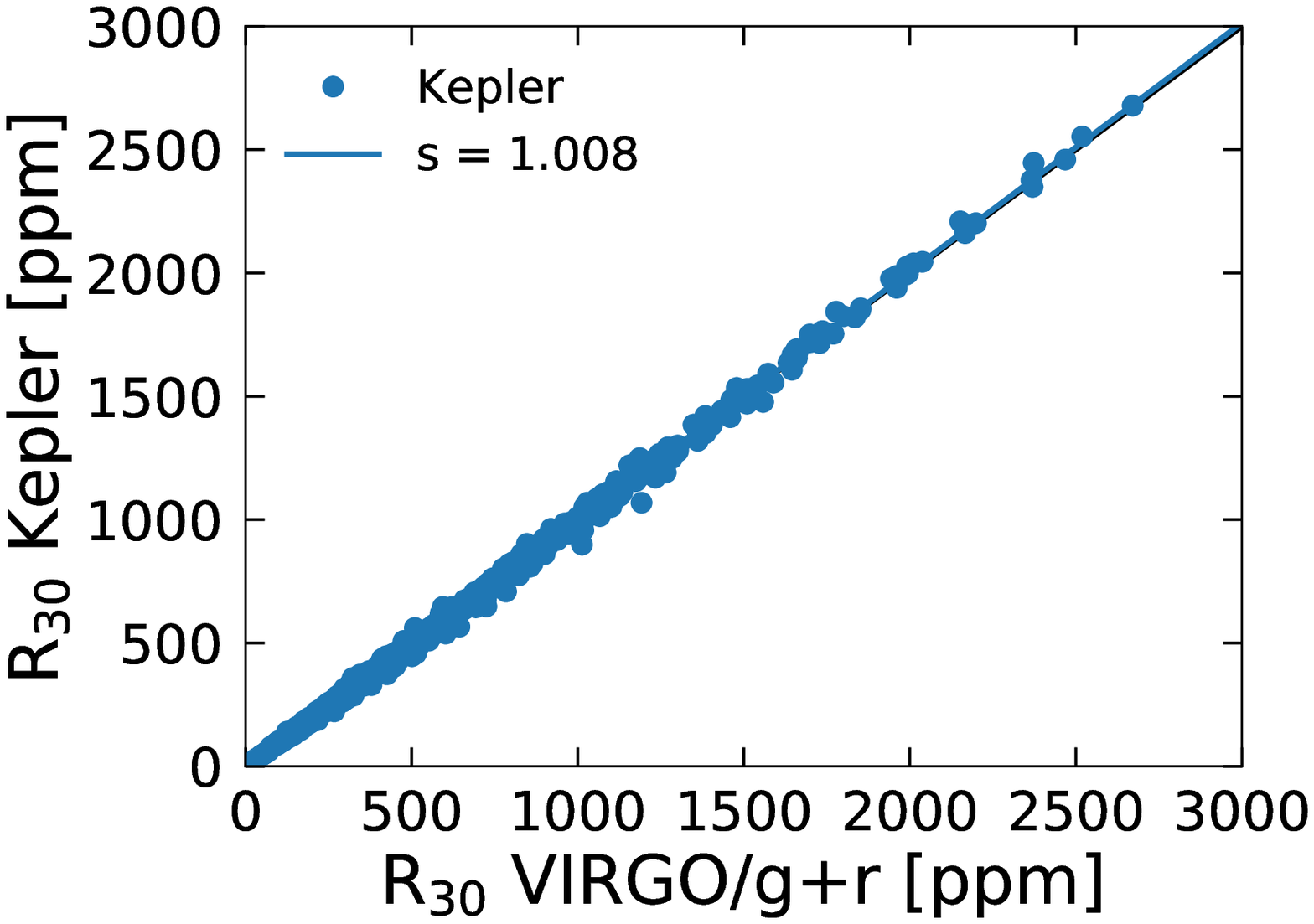}
\caption{Linear regression between the $R_{30}$ calculated for  VIRGO/g+r and for \textit{Kepler} light curves. The black solid line represents a linear regression with slope $=$ 1.}
\label{fig:K_V_regression}
\end{figure}

Several studies \citep[see e.g.][]{Basri2010,Harrison2012} have assumed that the amplitude of the rotational solar variability as it would be measured by Kepler is very close to the amplitude calculated for the combined green and red VIRGO/SPM light curves (in the following VIRGO/g+r). %  as a representative of the solar variability in the \textit{Kepler} passband.
Here we test this hypothesis.  %investigate, how the variability in the composite VIRGO/g+r band compares to the Sun as it would be observed by \textit{Kepler}.
The variability $R_{30}$ for \textit{Kepler} in comparison to  VIRGO/g+r is shown in Fig.~\ref{fig:K_V_LC}, which is limited to the same time interval as Fig.~\ref{fig:filters_R_30}. One can see that both curves are remarkably similar to one another.  To test the similarity quantitatively, we show the linear regression of $R_{30}$ between {\it Kepler} and VIRGO/g+r for four solar cycles (21-24) in Fig.~\ref{fig:K_V_regression}. The  Pearson correlation coefficient is very high (0.999) and the slope deviates by just  +0.8\% of unity averaged over four solar cycles.  While these calculations relate to the amplitude of the rotational variability, $R_{30}$,  we will additionally calculate regressions between {\it Kepler} and Virgo light curves in Sect.~\ref{Kepler-VIRGO}. We also directly connect the TSI and VIRGO/g+r rotational variability. The linear regression between $R_{30}$ in TSI and VIRGO/g+r  results in a slope of 0.88 ($\pm$ 0.002) and a Pearson correlation coefficient of 0.995.

%We can now use these numbers and  Table \ref{tab:regressions} to estimate the difference between VIRGO/g+r and the \textit{Kepler} filter. Using the value from Table \ref{tab:regressions} of a slope of 1.189 (corresponding to a difference of $+19$\%) and the difference of 11.6\% between the TSI and VIRGO/g+r, we find that the difference between \textit{Kepler} and VIRGO/g+r is about 6\%.
%VIRGO/g+r and the TESS filter. Using the value from Table \ref{tab:regressions} of a slope of 0.971 (corresponding to a difference of $-3$\%) and the difference of 11.6\% betwee%VIRGO/g+r and the TESS filter. Using the value from Table \ref{tab:regressions} of a slope of 0.971 (corresponding to a difference of $-3$\%) and the difference of 11.6\% between the TSI and VIRGO/g+r, we find that the difference between TESS and VIRGO/g+r is 9\%.

\section{Correction for the inclination}\label{Inclination}
\subsection{Approach}

The results presented in Sect. \ref{results_eq} are for the Sun viewed from the ecliptic plane and apply to stars that are viewed approximately equator-on. However, this is not always the case and often the inclination of a star is unknown.
Calculations of the solar variability as it would be measured by an  out-of-ecliptic observer  demand information about the distribution of magnetic features on the far-side (for the Earth bound observer) of the Sun. N20 have used a surface flux transport model (SFTM) to obtain the distribution of magnetic features of the solar surface, which was then fed into the  SATIRE-model to calculate solar brightness variations as they would be seen at different inclinations. 

The SFTM is an advective-diffusive model for the passive transport of the radial magnetic field on the surface of a star, under the effects of large-scale surface flows. In this model, magnetic flux emerges on the stellar surface in the form of bipolar magnetic regions (BMRs). We employ the SFTM in the form given by \cite{Cameron2010} and follow the approach of N20 to simulate light curves of the Sun at different inclinations and with various filter systems. 
The emergence times, positions and sizes of active regions in our calculations are determined using the semi-empirical sunspot-group record produced by \cite{Jiang2011_1}. This synthetic record was constructed to represent statistical properties of the  Royal Greenwich Observatory sunspot record. We additionally randomised the longitudes of the active-region emergences in the \cite{Jiang2011_1} records. Such a randomisation is needed to ensure that the near- and far-side of the Sun have on average equal activity, which is a necessary condition for reliable calculations of the inclination effect. As a result, our calculations reproduce the statistical properties of a given solar cycle but they do not represent  the actual observed BMR emergences for that specific cycle. We stress that in N20 we developed the model outlined above to study the effect of the inclination on the power spectra of solar brightness variations. Here we use this model to study explicitly how the amplitude of the variability on the rotational timescale depends on the inclination in different filters.

\subsection{Results}

In the following, we displace the observer out of the solar equator towards the solar north pole. This corresponds to inclinations below 90$^{\circ}$ . We quantify the rotational variability using the $R_{30}$ metric introduced in the previous section. To represent an average level of solar activity, we limit the analysis to cycle 23, which was a cycle of moderate strength.
 
  \begin{figure*}%[hbt!]
\centering
\includegraphics[width=\textwidth]{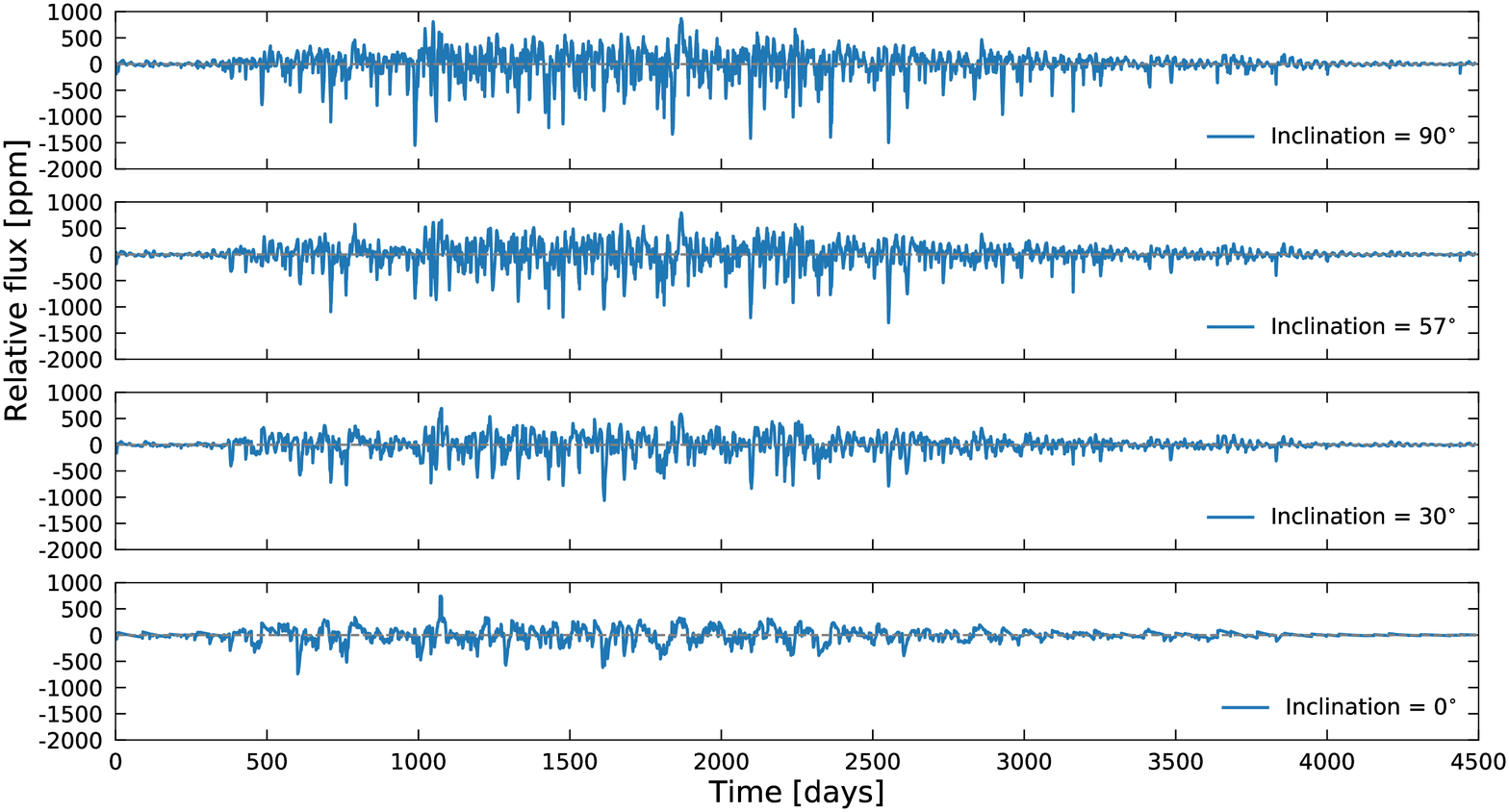}
\caption{Synthetic solar light curves covering solar cycle 23 in the \textit{Kepler} passband as it would appear at different inclinations.}
\label{fig:incl_LC}
\end{figure*}

  \begin{figure*}%[hbt!]
\centering
\includegraphics[width=\textwidth]{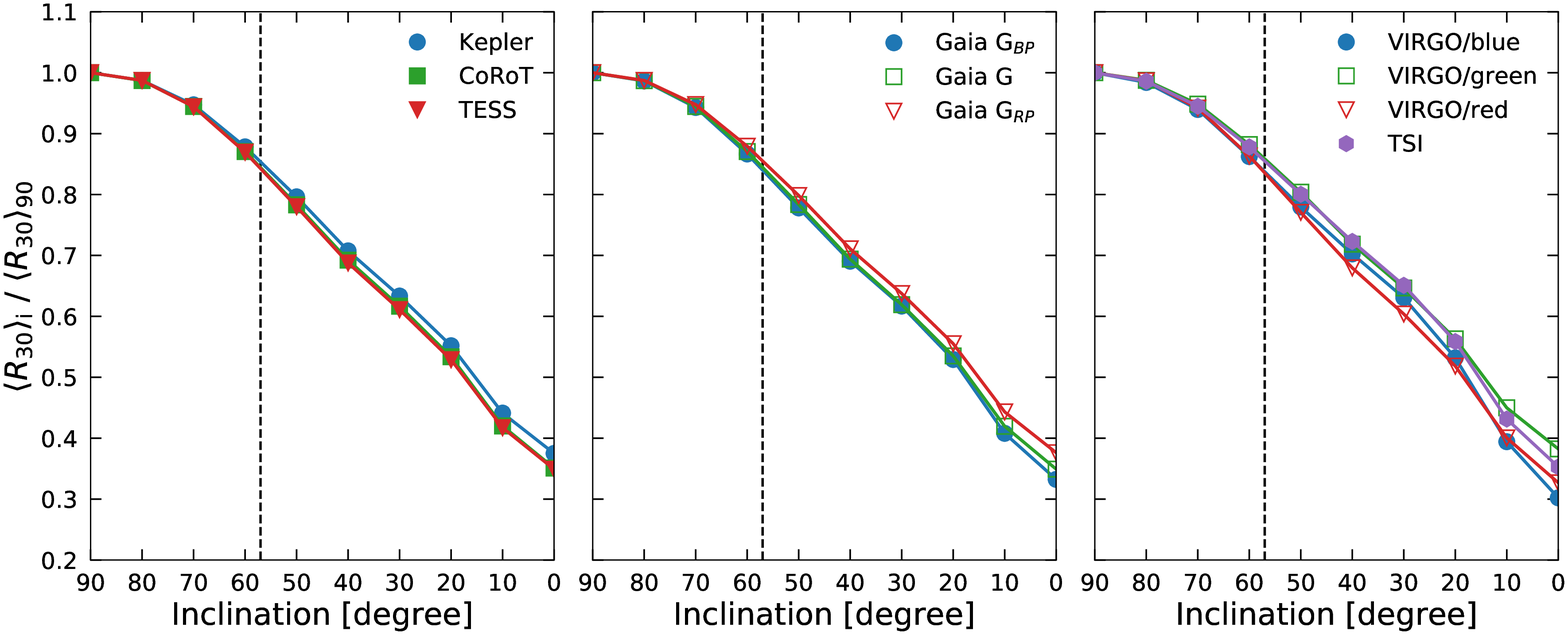}
\caption{Dependence of the mean variability in $R_{30}$ on the inclination (termed $\langle R_{30}\rangle_{\rm i}$, where $i$ stands for the inclination). All values have been normalised to the respective equator-on ($i=90^{\circ}$) value, here called $\langle R_{30}\rangle_{\rm 90}$. Individual curves represent different filter systems. Left panel: \textit{Kepler}, TESS and CoRoT, middle panel: the three \textit{Gaia} filters, right panel: the three VIRGO filters and the TSI. The vertical dashed black lines indicates an inclination of 57$^{\circ}$.}
\label{fig:ratios_inclination}
\end{figure*}

We show the calculated solar light curve as it would be observed by \textit{Kepler} at various inclinations in Fig.~\ref{fig:incl_LC}. For this, we divided the time-series for cycle 23 in 90-day segments, which correspond to \textit{Kepler} quarters. Within each quarter, we de-trended the light curves. 90$^{\circ}$ corresponds to an ecliptic-bound observer, 57$^{\circ}$  represents a weighted mean value of the inclination with weights equal to the probability of observing a given inclination ($sin(\mathrm{i})$) for the inclination i), and 0$^{\circ}$ corresponds to an observer facing the north pole. We additionally show 30$^{\circ}$ as an intermediate point between 57$^{\circ}$ and 0$^{\circ}$. For the inclination values of 30$^{\circ}$, 57$^{\circ}$, and 90$^{\circ}$ the variability is brought about by the solar rotation, as well as the emergence and evolution of magnetic features. 
A polar-bound observer does not observe the rotational modulation, as there is no transit of magnetic features and the variability is merely generated by their emergence and evolution \citep[see][for further details]{Nina2020}. We emphasise just the reduction in the amplitude of the variability with decreasing inclination, but also the change in the shape of the light curve. This is particularly visible in a comparison of the top and the bottom panels of Fig.~\ref{fig:incl_LC}. 

In Fig.~\ref{fig:ratios_inclination}, we show how  $R_{30}$ averaged over cycle 23 would change as we displace the observer out of the ecliptic plane. To ease the comparison, we normalised each value of the cycle-averaged variability, $\langle R_{30}\rangle_{\rm i}$, to the corresponding value for the ecliptic view, $\langle R_{30}\rangle_{\rm 90}$. Figure~\ref{fig:ratios_inclination} shows that the rotational variability decreases monotonically with decreasing inclination.
This trend is seen across all considered filter systems.
The differences in the inclination effect among the filter systems are due to the different dependencies of the facular and spot contrasts (as well as their centre-to-limb variations) on the wavelengths.

To evaluate the averaged effect of the inclination, we introduce a new measure that we call $\langle R_{30}\rangle$  and define as
\begin{equation}
\langle R_{30}\rangle =\frac{\sum_{i}\langle R_{30}\rangle_{i}\cdot \mathrm{sin}(i)}{\sum_{i}\mathrm{sin}(i)},
\end{equation}
\noindent where $i$ is the inclination. The factor $\mathrm{sin}(i)$ ensures that the corresponding values of $\langle R_{30}\rangle_{i}$ are weighted according to the probability that a star is observed at inclination $i$. The $\langle R_{30} \rangle$ value represents the variability of the Sun averaged over all possible inclinations. In other words, if we observed many stars analogous to the Sun with  random orientations of the rotation axes, their mean variability would be given by the $\langle R_{30}\rangle$ value. Therefore $\langle R_{30}\rangle$ should be used for the solar-stellar comparison rather than the $\langle R_{30}\rangle_{90}$ value.

We present $\langle R_{30}\rangle$ normalised to  $\langle R_{30}\rangle_{90}$ for all considered filter systems as well as the $\langle R_{30}\rangle$ values themselves in Table~\ref{tab:inclination}. In the second column of Table~\ref{tab:inclination} we give the inclination-corrected value of the mean rotational variability from Table~\ref{tab:mean_R_30} for easier application of our results. On average, all filter systems show 15\% less variability compared to the equatorial case.
This implies that the slopes of the linear regressions between the $R_{30}$ values in different passbands  and the TSI have to be corrected for the inclination. When comparing stellar measurements in the \textit{Kepler} passband with the TSI records this means the following: if stars are observed from their equatorial planes, the variability in \textit{Kepler} is about 12\% higher than in the TSI (see the slope given in Table \ref{tab:regressions}). However, when comparing the Sun to a group of stars with random orientations of rotation axes the inclination effect must be taken into account as well. It will reduce stellar variability observed in \textit{Kepler} passband by approximately 15\%. Coincidentally, both effects almost exactly cancel each other and the observed TSI variability appears to be a very good metric for the solar-stellar comparison of \textit{Kepler} stars in a statistical sense. 
\begin{table}%[!htbp] 
\begin{center}
\begin{tabular}{lcc} 
\hline \hline%
	&$\langle R_{30}\rangle$/$\langle R_{30}\rangle_{90}$ [\%]	&$\langle R_{30}\rangle$	[mmag]	\\
\hline
TSI                 	&	85.1	&	0.58	\\ %showing the TSI here probably does not make much sense
\textit{Kepler}	        &	84.8	&	0.62	\\
CoRoT	                &	83.8	&	0.61	\\
TESS	                &	84.0	&	0.56	\\
\textit{Gaia G$_{BP}$}	&	83.7	&	0.72	\\
\textit{Gaia G}	        &	84.1	&	0.62	\\
\textit{Gaia G$_{RP}$}	&	84.9	&	0.52	\\
VIRGO/blue	            &	83.9	&	0.98	\\
VIRGO/green	            &   85.3	&	0.76	\\
VIRGO/red	            &	83.4	&	0.50	\\
\hline \hline
\end{tabular}
\end{center}
\caption{$\langle R_{30}\rangle$/$\langle R_{30}\rangle_{90}$    and      $\langle R_{30}\rangle$ values for different filter systems. For $\langle R_{30}\rangle$ we multiplied $\langle R_{30}\rangle$/$\langle R_{30}\rangle_{90}$ by the corresponding value for $\langle R_{30}\rangle_{90}$ from Table \ref{tab:mean_R_30}.  Time averaging is performed over the solar cycle 23.}
\label{tab:inclination}
\end{table}

We have shown in Sect.~\ref{Equator} that the amplitude of solar rotational variability as it would be measured by \textit{Kepler} can be very accurately approximated by calculating the amplitude of the VIRGO/g+r light curve. However, when comparing brightness variations of the Sun to those of a large group of stars with unknown inclinations, one should use solar variability averaged over all possible inclinations rather than solar variability observed from the ecliptic plane.
We have established  that the effect of a random inclination decreases the variability in the \textit{Kepler} passband by 15\%. Taking this into account, the relative difference between the variability in VIRGO/g+r and the solar variability in \textit{Kepler}  averaged over inclinations is $-14$\%. Unlike for the TSI, corrections for the passband and the inclination only partly compensate each other. We therefore suggest that the TSI is a better representative of the Sun as it would be observed by \textit{Kepler} than VIRGO/g+r, if the inclination of a star is unknown.

%\section{VIRGO/SPM versus \textit{Kepler} variability}\label{Kepler-VIRGO}
\section{Modelling \textit{Kepler} light curves using VIRGO/SPM}\label{Kepler-VIRGO}

The the previous sections, we have quantitatively validated the argument of \cite{Basri2010} that the VIRGO/g+r light curve correspond to the same variability as the \textit{Kepler}  light curve if both light curves are recorded from the solar equatorial plane. 
In this section we perform complementary calculations.
Namely, we test if the  \textit{Kepler} light curve can be modelled as a linear combination of solar light curves in the different VIRGO/SPM channels. We restrict our calculations to solar cycle 23. All light curves are computed with the N20 model.

We divide all light curves into 90-day segments and calculate the relative flux within these segments (i.e. we consider the same normalisation of light curves as shown in Figs. \ref{fig:filters_LC} and \ref{fig:incl_LC}). Next, we apply multiple linear regression to fit the \textit{Kepler} light curve with the VIRGO/SPM/green+red light curves, to find the best set of coefficients for the linear fit.

 \begin{figure*}[hbt!]
\centering
\includegraphics[width=\textwidth]{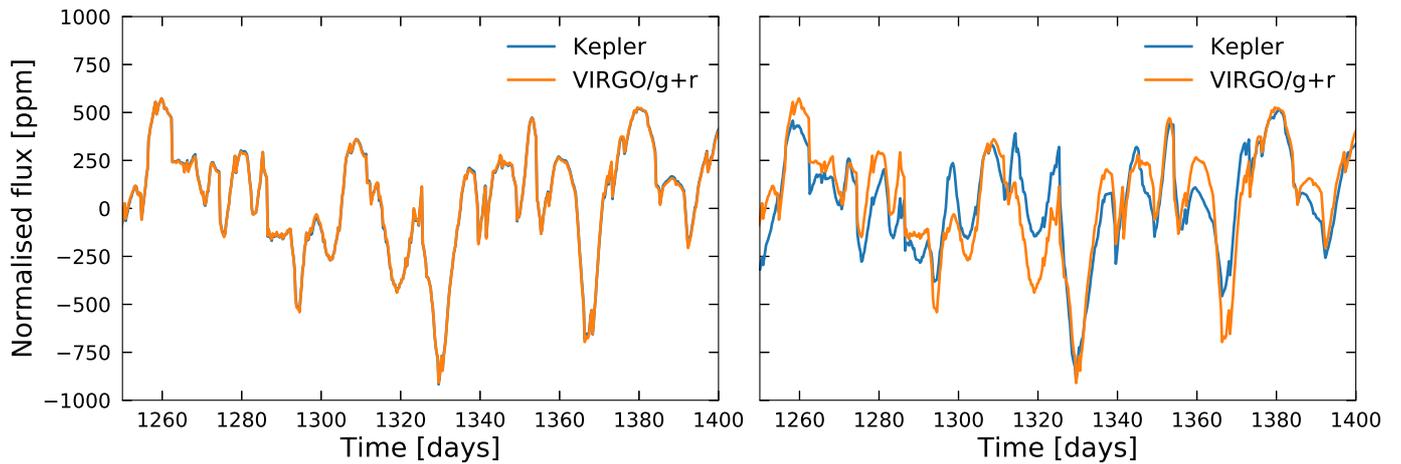}
\caption{Comparison between the \textit{Kepler} and the regressed VIRGO/green+red light curve. Left:i = 90$^{\circ}$,  right:i=57$^{\circ}$. For the coefficients of the fit, see text.} 
\label{fig:fit}
\end{figure*}

For an ecliptic-bound observer we write the multiple linear regression in the form
\begin{equation}
    K_{i,\,SPM} =   a\cdot V_g+ b\cdot V_r ,
\label{eq:fit_2_90}
\end{equation}
where $i$ is the inclination, $V_g$ and $V_r$ are solar light curves  in VIRGO/SPM blue green and red filters (corresponding to the equatorial plane). The best fit for $i=90^{\circ}$ yields  a$=$0.275 ($\pm$ 0.001) and b$=$0.619 ($\pm$ 0.002). The $r^2$-value is 0.999. Such a high correlation is not surprising giving the similarity between the VIRGO/SPM and \textit{Kepler} rotational variability discussed in Sect.~\ref{Equator}.
Next, we apply the multiple-regression model to simulate the out-of-ecliptic \textit{Kepler}-like light curve using the light curves in the SPM channels as the input.
Given that an inclination of 57$^{\circ}$ is often used to represent a statistical mean of all possible inclination values, we fit the \textit{Kepler} light curve observed at $i=57^{\circ}$ with a linear combination of VIRGO light curves (observed at $i=90^{\circ}$). The best fit results in the following coefficients: a$=-$0.421 ($\pm$ 0.018) and b$=$1.418 ($\pm$ 0.027), with $r^2 = 0.83$.

%We again start with a linear combination of the green and red light curves.
%The fit yields
%\begin{equation}
%    K_{57,\,g+r} =  1.355 (\pm 0.028)\cdot V_g - 0.3613(\pm 0.019)\cdot V_r.
%\label{eq:fit_2_57}
%\end{equation}
%The Pearson correlation coefficient is only 0.831.
%\cnk{Why "only"? In general, this is a high Rc. But you can only say if it is high or not if you do a significance test.}
%By adding the blue channel we obtain
%\begin{equation}
%    K_{57,\,b+r+g} = 0.135 (\pm 0.011) \cdot V_b + 1.087(\pm 0.035)\cdot V_g  - 0.347 (\pm 0.019)\cdot V_r.
%\label{eq:fit_3_57}
%\end{equation}
%The correlation coefficient stays low at 0.832. 

%We show the comparison between the \textit{Kepler} and all three combined VIRGO filter light curves in Fig.~\ref{fig:fit}.
Figure~\ref{fig:fit} compares the Kepler-like light curve with the regression model using light curves in all three VIRGO filters for $i=90^{\circ}$ and $i=57^{\circ}$.
For the 90$^{\circ}$ inclination case (left panel in Fig.~\ref{fig:fit}), the differences between the two light curves are basically invisible, but for the 57$^{\circ}$ case, the differences are quite pronounced. These differences have various origins. In particular, the transits of magnetic features would take different times for the ecliptic and out-of-ecliptic observer. Furthermore, with decreasing inclination, the facular contribution becomes stronger, while the spot contribution weakens \citep[see e.g.,][]{Shapiro2016}, changing the shape of the light curve. It is therefore necessary to take the actual distribution of magnetic features into account when modelling light curves observed at different inclination angles.

\section{Conclusions}\label{Conclusion}
In this study we give recipes how to treat two problems hindering  the comparison between solar and stellar rotational brightness variations: the difference between the spectral passbands utilised for solar and stellar observations as well as the effect of inclination.

To quantify the effect of different spectral passbands on the rotational variability represented through the $R_{30}$ metric, we employed the SATIRE-S model. We found that the rotational variability observed through the filter systems used by the \textit{Kepler} and CoRoT missions is around 12\% higher than the TSI, whereas the variability in the TESS passband is about 7\% lower. For \textit{Gaia G} we find +15\% and for \textit{Gaia G$_{BP}$} +30\% difference in the amplitude of the rotational variability compared to the TSI, whereas \textit{Gaia G$_{RP}$} shows a difference of -7\%. These numbers are valid for equator-on observations on rotational timescales.

 Previous studies have used combinations of the green and VIRGO red light curves for solar-stellar comparisons \citep[see e.g.][]{Basri2010,Gilliland2011,Harrison2012}. 
 We used linear regressions of the rotational variability between the two combined VIRGO/SPM passbands and the solar variability as \textit{Kepler} would observe it, to test the goodness of that comparison. We find that the variability in \textit{Kepler} is 7\% higher compared to that of VIRGO/green+red. Moreover, we found that  the sum of the VIRGO green and red light curves very accurately represents the solar light curve in the \textit{Kepler} passband. This is valid, however, only valid for the Sun observed from the ecliptic. We show that a linear combination of VIRGO/SPM passbands cannot accurately reproduce the solar \textit{Kepler} light curve observed out-of-ecliptic.
 
 We have calculated the dependence of the rotational variability on inclination by following the approach in \cite{Nina2020}. In this approach, an SFTM is used to simulate the distribution of magnetic features on the surface of the Sun, which is then used to compute the brightness variations with SATIRE. We find that across all filter systems discussed in this study the rotational variability drops by about 15\% when it is averaged over all possible directions of the rotation axis. Given that the \textit{Kepler} rotational variability as observed from the ecliptic plane is 12\% higher than the TSI rotational variability, we conclude, that the TSI is the best proxy for the solar rotational brightness variations if they would be observed by \textit{Kepler} when the inclination effect is considered.

%\cnk{Can you give a recommendation? Should TSI be used instead? Does it make sense to consider SPM then? I mean not in this work - but as a recommendation. If TSI works better, is more stable and does not require any regression. IS not it worth of emphasising?}

%Using the sum of the green and red VIRGO/SPM passband is a good representative of the solar solar rotational variability as \textit{Kepler} would observe it, if a given star is also observed almost equator-on. 

\begin{acknowledgements}
We thank Chi-Ju Wu, whose master thesis has sparked the idea for the manuscript.
The research leading to this paper has received funding from the European Research Council under the European Union’s Horizon 2020 research and innovation program (grant agreement No. 715947). 
SKS acknowledges financial support  from the BK21 plus program through the National Research Foundation (NRF) funded by the Ministry of Education of Korea.
\end{acknowledgements}

\bibliographystyle{aa}
\bibliography{bib}

\end{document}